\documentclass[preprint,authoryear,12pt]{elsarticle}

\usepackage{epsfig}

\usepackage{amssymb}

\usepackage[ps2pdf,%
a4paper=true,%
breaklinks=true,%
colorlinks=true,%
pdfauthor={First Author et al.},%
pdftitle={Template for manuscripts in Advances in Space Research}%
]{hyperref}

\journal{Advances in Space Research}

\begin{document}

%%%%%%%%%%%%%%%%%%%%%%%%%%%%%%%%%%%%%%%%%%%%%%%%%%%%%%%%%%%%%%%%%%%%%%%%%%%%%
%% Frontmatter
\begin{frontmatter}


% Use the tnoteref command within \title and fnref within \author or \address for footnotes;
% use the corref command within \author for corresponding author footnotes;
% \author{Name\corref{cor1}\fnref{label2}}
% \address{Address\fnref{label3}}The Sloan Digital Sky Survey Photometric System

\title{Lunar Exosphere Influence on Lunar-based Near-ultraviolet Astronomical Observations \tnoteref{footnote1}}
\tnotetext[footnote1]{This template can be used for all publications in Advances in Space Research.}

% Use optional labels to link authors explicitly to addresses:

\author{J. Wang\corref{cor}}%\fnref{footnote2}}
\address{National Astronomical Observatories, Chinese Academy of Sciences,
20A, Datun Road, Chaoyang District, Beijing, China, 100012}
\cortext[cor]{Corresponding author}
%\fntext[footnote2]{Additional information regarding the corresponding author}
\ead{wj@bao.ac.cn}

% Url can be given like this:
% \ead[url]{http://www.elsevier.com/wps/find/authorsview.authors/latex}
\author{J. S. Deng}
\address{National Astronomical Observatories, Chinese Academy of Sciences,
20A, Datun Road, Chaoyang District, Beijing, China, 100012}
\ead{jsdeng@bao.ac.cn}
\author{J. Cui}
\address{National Astronomical Observatories, Chinese Academy of Sciences,
20A, Datun Road, Chaoyang District, Beijing, China, 100012}
\ead{cuij@nao.cas.cn}
\author{L. Cao}
\address{National Astronomical Observatories, Chinese Academy of Sciences,
20A, Datun Road, Chaoyang District, Beijing, China, 100012}
\ead{caoli@bao.ac.cn}
\author{Y. L. Qiu}
\address{National Astronomical Observatories, Chinese Academy of Sciences,
20A, Datun Road, Chaoyang District, Beijing, China, 100012}
\ead{qiuyl@bao.ac.cn}
\author{J. Y. Wei}
\address{National Astronomical Observatories, Chinese Academy of Sciences,
20A, Datun Road, Chaoyang District, Beijing, China, 100012}
\ead{wjy@bao.ac.cn}

%\fntext[footnote3]{Additional information about the second and third authors}
%\ead{more@email.addresses}

%\author{More Authors\fnref{footnote4}}
%\address{Address of the co-authors}
%\fntext[footnote4]{Additional information about the co-authors}
%\ead{more@email.addresses}

\begin{abstract}
%The near-ultraviolet sky background emission from the solar-forced lunar exosphere is predicted basing upon the
%current knowledge of the exosphere. The emission is calculated by a sum of the contributions from
%Rayleigh scattering, resonance fluorescence and emissive photodissociation mechanisms.

The potential effect of the lunar exosphere on the near-ultraviolet sky background emission
is predicted for Lunar-based Ultraviolet Telescope (LUT: a funded Chinese scientific payload for the Chang'e-III mission).
Using the upper limit on the OH concentration inferred from the recent MIP CHACE results,
our calculations show that the sky brightness due to the illuminated exosphere is
$<8.7\ \mathrm{photon\ s^{-1}\ cm^{-2}\
arcsec^{-2}}$ within the wavelength range 245-340 nm.
By evaluating the signal-to-noise ratios of observations of an AB=13~mag point source at
a series of sky background levels,
our analysis indicates that the detection performance of LUT can be moderately
degraded by the lunar exosphere emission in most cases. An AB=13~mag point source can
still be detected by the telescope at a signal-to-noise ratio more than 8 when the
OH concentration is less than $2\times10^8\ \mathrm{molecules\ cm^{-3}}$. However, the
effect on the performance is considerable when the exosphere is as dense as suggested by CHACE.
\end{abstract}


\end{frontmatter}

\parindent=0.5 cm

%%%%%%%%%%%%%%%%%%%%%%%%%%%%%%%%%%%%%%%%%%%%%%%%%%%%%%%%%%%%%%%%%%%%%%%%%%%%%
%% Main text
\section{Introduction}

Lunar-based astronomical observations have advantages over both ground-based and
space-based observations. 1) It has been known for decades that the Moon is surrounded by an extremely tenuous atmosphere
($\sim10^{4-5}\ \mathrm{molecules\ cm^{-3}}$ at night, and a much higher value
$\sim10^{7-8}\ \mathrm{molecules\ cm^{-3}}$ in the daytime, see Heiken et al. 1991; Stern 1999 for a review).
This means that atmospheric opacity, atmospheric scattering/emission and, the atmospheric turbulence are absent.
2) Unlike space-based observations, the Moon provides a large stable platform for maintaining
astronomical instruments in permanently stable configurations. 3) The sky's diurnal motion on the
Moon is $0.\symbol{125}55\ \mathrm{s^{-1}}$, 27 times slower than on the Earth.
This allows long term monitoring for 13 days without interruption.
4) The temperature in the permanent shadow regions (PDRs) at both poles of the Moon
could permanently be as low as 30K. PDRs are therefore ideal places for infrared observations.
5) Observations at very low frequency ($<$10MHz, VLF) are feasible on the Moon, but not on Earth,
because  VLF electromagnetic waves cannot penetrate the Earth's ionosphere.

The Apollo-16 mission performed far-ultraviolet observations on
the lunar surface in 1972 for the first time in history (Page \&
Carruthers 1977; Carruthers \& Page 1977; Carruthers \& Page
1972). The observations were carried out with a 3-inch Schmidt
telescope equipped with a far-ultraviolet camera/spectrograph
operating in the wavelength range from 100-160 nm. The
field-of-view was about 20\symbol{23}, and the limiting magnitude
was 11. The telescope was placed in the shadow of the Lunar Module
to avoid heating by the Sun. In total, 178 images were obtained by
the telescope, and delivered to the Earth by the astronauts. In
addition to the Earth's geocorona, the observed objects included
star clusters and the Magellanic clouds. These observations,
however, contributed little to astronomical knowledge because of
the low level of the technology used at that time.

The Chinese Chang'e-III mission is designed to deploy a dedicated, robotic Lunar-based Ultraviolet Telescope
(LUT, Cao et al. 2011) on the surface of the Moon. LUT will perform astronomical observations in
near-ultraviolet (NUV) band. The main scientific goal of LUT is to monitor variable stars and active galaxies in
the NUV band for more than a dozen days.
Objects with large brightness variations in the NUV band include cataclysmic variable stars,
large/ small mass binaries and novae, quasars and Blazars, dMe stars, and Lyr RR stars.
The variations of the stellar temperature, radius, and
accretion rate of these compact objects can be studied, useful to verify current stellar atmosphere models
and to investigate the origin of the instability.
LUT will additionally perform low-Galactic-latitude sky surveys as a complement to the NASA
GALEX (Galaxy Evolution Explorer) mission (Martin et al. 2005) whose onboard MCPs prevent from observing the bright
objects with NUV$<$10mag (Morrissey et al. 2005).
At present, lunar-based observations are only feasible in the Moon daytime mainly due to
the lack of electronic power supply at night.

The Indian Chandrayaan-1 payload
Chandra's Altitude Composition Explore (CHACE) recently claimed a lunar exosphere pressure of
$10^{-7}$torr (Sridharan et al. 2010a, b), higher than that found by the Apollo missions by 2 orders
of magnitude (see Section 3 for more details). Because of these new results,
this paper wishes to estimate the effect of this potentially enhanced lunar exosphere on the signal-detection performance of LUT.
The sunlight
can be scattered and/or re-radiated by the particles of the lunar exosphere.
These processes mainly include Rayleigh scattering, resonant scattering, and emissive photodissociation.

The paper is organized as follows. After describing the basic properties of LUT in Section 2,
Section 3 briefly summarizes previous measurements of the lunar exosphere.
The sky background brightness is calculated for the lunar exosphere in Section 4.
The estimated brightness is then compared with the brightness of a point source with AB=13~mag.
Section 5 calculates the count rates
at the LUT focal plane, and estimates the final signal-to-noise ratio (S/N) for an observation of a point source.

\section{Lunar-based Ultraviolet Telescope}

As part of the scientific payload of the Chang'e-III mission, LUT will land on the Moon and
work in the lunar daytime (Cao et al. 2011).
A limiting magnitude of AB=13~mag is designed for an exposure time of 30 seconds.
The optical system of LUT is a F/3.75 Ritchey-Chr\`{e}tien telescope with an aperture of 150~mm.
A pointing flat mirror mounted on a two-dimensional gimbal is used to
point and track a given object. An ultraviolet-enhanced CCD E2V47-20, manufactured by the EEV Company,
is chosen as the detector mounted at the Nasmyth focus. Table 1 tabulates the designed performance
parameters on which our subsequent calculations are based. We refer the readers to Cao et al. (2011)
for more details on the mission's concept and design.

\begin{table}
\small
\caption{The designed performance parameters of LUT}
\begin{tabular}{lccc}
\hline
Parameter& symbol & unit& Value\\
\hline
Wavelength range &\dotfill & nm & 245-340 \\
Aperture size    & $d$ & cm & 15 \\
F-number              & \dotfill &\dotfill & 3.75\\
Pixel size of CCD & $d_p$ & $\mu$m &  13 \\
Average CCD QE  & $\overline{QE}$ & \dotfill &0.4 \\
Optical efficiency  & $\eta_{\mathrm{opt}}$ & \dotfill & 0.09 \\
Optics PSF & $f_p$ &\dotfill & 80\% energy within $3\times3$ pixels \\
CCD Readout noise & $RN$ & $\mathrm{e^-\ pixel^{-1}\ read^{-1}}$ & 8\\
Dark current & $D$ & $\mathrm{e^-\ s^{-1}\ pixel^{-1}}$ & 1.0 (temperature$<$-20)\\
CCD Gain &  $G$ & $\mathrm{e^-\ ADU^{-1}}$ & 1\\

\hline
\end{tabular}
\label{table1}
\end{table}

\section{Measurements on Lunar Exosphere}

Our current understanding of the lunar exosphere is mainly provided by the measurements done
by the Apollo missions in
1970s (Heiken et al. 1991). The Apollo-12, -14 and -15 missions deployed three Cold Cathode Gage Experiments (CCGEs) on the
lunar surface to record the gas concentration of the exosphere at the landing sites.
Although the Apollo-12 instrument failed after less than one day of operation, the other two instruments
operated until the mid-1970s. CCGEs accurately recorded a gas concentration of
$\sim10^{4-5}\ \mathrm{molecules\ cm^{-3}}$ at night. The daytime data are, however, hard to
interpret because of the outgassing from the
instruments and the limited dynamic ranges of the instruments. Briefly, CCGEs
recorded a fast rise (decrease) in gas concentration at each dawn (dusk), and determined an uncertain upper limit of
$\sim10^{7}\ \mathrm{molecules\ cm^{-3}}$ in the daytime since the instruments were saturated
soon after sunrise (Johnson et al. 1972).

The final Apollo mission, Apollo 17, deployed a mass spectrometer, Lunar Atmosphere Composition
Experiment (LACE), on the lunar surface to determine the abundance of the exospheric gas (Hodges, 1973;
Hodges et al. 1972).
The mass range of LACE is 1-110 amu, and the sensitivity is $1\ \mathrm{counts\ s^{-1}}$
(corresponding to $200\ \mathrm{molecules\ cm^{-3}}$). The nine-months of operation of LACE
shows that the lunar exosphere is mainly composed of $\mathrm{^{20}Ne}$, He, $\mathrm{H_2}$,
$\mathrm{^{40}Ar}$, and $\mathrm{CO_2}$.

Knowledge about the lunar exosphere is enhanced by the recent on-orbit measurements done by
a mass spectrometer, Chandra's Altitude Composition Explore (CHACE), onboard the Moon
Impact Probe (MIP) of the Indian Chandrayaan-1 mission (Sridharan et al. 2010a, b).
CHACE is sensitive to a mass range of 1-100~amu. The partial pressure sensitivity is
$\sim10^{-13}$~torr, significantly lower than the
sensitivity of LACE by four orders of magnitude. CHACE sampled the exosphere
gas at the sunlit side of the Moon every four seconds,
after being released from the stationary orbit at an altitude of about 98~km.
The sampling resolution is 0.\symbol{23}1 in latitude, and 250~m in altitude.
The measurements indicate that the total pressure of the exosphere is about $10^{-7}$ torr,
two orders of magnitude higher than the values previously reported by the Apollo missions.
In contrast to the results from the Apollo-17 mission, the exosphere is found to be dominated
by $\mathrm{H_2O}$ and $\mathrm{CO_2}$ molecules. In addition, CHACE detected a pressure
increase of a factor of about two when the instrument raced towards the surface at the south
pole from the release point.

Outgassing from the CHACE instrument may have added to these results.
Therefore, the pressures reported by CHACE should be considered as upper limits for
our subsequent calculations.

\section{Emission from Point Source and Sky Background}

In this section, we calculate the sky background brightness produced
by the lunar exosphere, and compare the predicted brightness
with the brightness of a point source with an AB magnitude of 13~mag.

\subsection{Point source}

The broadband AB magnitude of an astronomical point-source is defined as (Fukugita et al. 1996)
  \begin{equation}
    \label{eq:1}
 {%
 \mathrm{mag_{AB}}=-2.5\log\frac{\int f_\nu S_\nu d\ln\nu}{\int S_\nu d\ln\nu}-48.6
    }
%  \overfullrule 5pt
%  \mathindent\linewidth\relax
%  \advance\mathindent-259pt
  \end{equation}
where $f_\nu$ is the specific flux density of the object in unit of
$\mathrm{erg\ s^{-1}\ cm^{-2}\ Hz^{-1}}$, and $S_\nu$ is the total efficiency of
the photometric system at frequency $\nu$. The total efficiency $S_\nu$ is obtained
from the product of the optical efficiency of the telescope,
the transparency of the filter used and the quantum efficiency of the detector.
The performance parameters listed in Table 1 yield an efficiency of
$\overline{S_\nu}=\eta_{\mathrm{opt}}f_p\overline{QE}=0.0288$ for LUT. With this
efficiency, the total photon flux $N_*$ within the wavelength range
from $\lambda_1$ to $\lambda_2$ can be determined from the magnitude by the formula
$N_*=1/h\ln(\lambda_2/\lambda_1)\times10^{-0.4(\mathrm{mag_{AB}}+48.6)}\
\mathrm{photon\ s^{-1}\ cm^{-2}}$, where $h=6.63\times10^{-27}\ \mathrm{erg\ s^{-1}}$ is
the Planck constant. For an point source with brightness AB=13~mag (i.e., the limiting magnitude of LUT),
the photon flux is predicted to be $N_*\approx11.4\ \mathrm{photon\ s^{-1}\ cm^{-2}}$
within the wavelength range from 245~nm to 340~nm.

\subsection{Sky background from the lunar exosphere}
\subsubsection{Rayleigh scattering}
Rayleigh scattering of sunlight by the molecules in the lunar exosphere contributes diffuse light to the sky background.
The scattered photons maintain their frequencies but change directions.
The differential cross-section for Rayleigh scattering strongly depends on the wavelength of
the incident photon $\lambda$ as
$d\sigma_R/d\Omega\propto(\lambda/a)^{-4}\sin^2\theta$, where $a$ is the length size of the scattering particle,
and $\theta$ is the scattering angle between the directions of the incident and scattered radiation.

Because the lunar exosphere is extremely tenuous, the monochromatic intensity of scattered light
along a line-of-sight can
be well calculated in the optically thin approximation as
$I=\int\eta_\mathrm{sc}dsd\lambda=\sec z\int\eta_\mathrm{sc}d\lambda dh$,
where $ds=\sec z dh$ is the element of path length.
Parameters $z$ and $h$ are the zenith angle and height above the surface,
respectively. Note that the expression $ds=\sec z dh$ is only valid when $z< 45\symbol{23}$.
For Rayleigh scattering, the emissivity
$\eta_{\mathrm{sc}}$ per unit volume can be evaluated as
$\eta_{\mathrm{sc}} =1/4\pi n\sigma_R F_\lambda$, where $n$ is the particle concentration in the
exosphere, $\sigma_R$ is the cross-section for Rayleigh scattering integrated over all directions, and
$F_\lambda$ is the specific solar flux density.

We assume that the vertical density distribution of each element in the exosphere can be described
by hydrostatic equilibrium $n=n_0 e^{-H/\Delta H}$, where $\Delta H$ is the scale height.
The brightness of the scattered sunshine can be predicted from the sum of the contributions of
all elements in the exosphere as follows
  \begin{equation}
    \label{eq:2}
 {I_p=\frac{\sec z}{4\pi hc}\sum\Delta Hn_0\int\sigma_R\lambda F_\lambda d\lambda
    }
  \end{equation}
where $h$ and $c=3\times10^{10}\ \mathrm{cm\ s^{-1}}$ is the Planck constant and light velocity,
respectively.

The gas abundances and corresponding scale heights used are tabulated in Table 2.
The values are taken from the Lunar Sourcebook (Heiken et al. 1991)
except for $\mathrm{H_2O}$ and $\mathrm{CO_2}$.
A mass spectrum taken by CHACE at an altitude $H\sim98$~km indicates that
the partial pressure of  $\mathrm{H_2O}$ molecule at that point is $p'\sim8\times10^{-8}$ torr
(Sridharan et al. 2010a, b). By assuming hydrostatic equilibrium,
the pressure at the surface is then inferred to be
$p=p'e^{H/\Delta H}\sim2.7\times10^{-5}$~Pa,
where the scale height is $\Delta H$=100~km.
We estimate the surface concentration of
$\mathrm{H_2O}$ molecules from its surface pressure by using the ideal gas law.
A collisionless exosphere model
(e.g., Chamberlain \& Hunten 1987) is used with the kinetic temperature rather than the thermal one.
The kinetic temperature $T$ can be derived from the scale height.
Because the gas in the exosphere is collisionless, the energy conservation of each particle results in a
relationship $\Delta H\sim kT/mg$,
where $m$ is the molecular mass, and $g$ is the gravitation at the surface.
A kinetic temperature $T\sim350$ K is therefore required for molecules to reach the scale height
$\Delta H$=100km.
Substituting this temperature into the ideal gas law, we derive an upper limit of
$n\sim6\times10^9\ \mathrm{molecules\ cm^{-3}}$ for the surface concentration of $\mathrm{H_2O}$ molecules.

An adjacent small peak at amu=17 can be identified in the CHACE mass spectrum.
The peak is likely produced by OH radicals (also perhaps by $\mathrm{NH_3}$).
Similarly as above, we obtain an upper limit of $n\sim2\times10^9\ \mathrm{molecules\ cm^{-3}}$
for the surface concentration of the OH radical.
There is another strong peak at amu=44 in the CHACE mass spectrum.
The same method yields an upper limit of $\sim1.3\times10^{10}\ \mathrm{molecules\ cm^{-3}}$ for
$\mathrm{CO_2}$ by assuming the same kinetic temperature
as for $\mathrm{H_2O}$.

\begin{table}
\small
\caption{Gas abundances and scale heights in the lunar exosphere.}
\begin{tabular}{lll}
\hline
Species &Surface concentration & Scale height\\
 & $\mathrm{cm^{-3}}$ & km \\
\hline
Ne  & $\sim10^4$ & 100 \\
He  & $\sim4.7\times10^3$ & 511 \\
$\mathrm{H_2}$ & $\sim9.9\times10^3$ & 1022 \\
Ar  & $\sim2\times10^3$ & 55\\
$\mathrm{H_2O}$ & $<6\times10^9$  & 100\\
OH & $<2\times10^9$  & 100\\
$\mathrm{CO_2}$ & $<1.3\times10^{10}$ & 46\\
\hline
\end{tabular}
\label{table1}
\end{table}

The cross-section for Rayleigh scattering is calculated as
$\sigma_R=a\lambda^{-4}(1+b\lambda^{-2}+c\lambda^{-4})$, where $\lambda$ is the wavelength
in units of \AA. The parameters $a$, $b$ and $c$ are listed in Table 3 for He, $\rm H_2$, Ar,
OH radical, $\mathrm{H_2O}$ and $\mathrm{CO_2}$.
The differential cross-section of Ne is adopted from the calculation based on the
Quantum defect theory\footnote{See http://adg.llnl.gov/Research/scattering/elastic.html.}.
Eq. (2) is then integrated within the wavelength range (i.e., 245nm-3400nm) by using the
1985 Wehrli Standard Extraterrestrial Solar Irradiance Spectrum\footnote{The
spectrum can be derived from http://rredc.nrel.gov/solar/spectra/am0/wehrli1985.new.html.}.
Our calculations yield a sky brightness at the zenith of
$I_p<1.5\times10^{-4}\ \mathrm{ photon\ cm^{-2}\ s^{-1}\ arcsec^{-2}}$,
which is 5 orders of magnitude lower than the brightness of a point-source with AB=13~mag.

\begin{table}
\caption{Rayleigh scattering parameters}
\begin{tabular}{lllll}
\hline
Species  & a & b & c & References\\
\hline
He  & $5.7\times10^{-14}$ & $0.4\times10^6$ & $0.18\times10^{12}$ & Behara et al. (2005)\\
$\mathrm{H_2}$ & $8.1\times10^{-13}$ & $1.5\times10^6$ & $1.98\times10^{12}$ & Dalgarno et al. (1962)\\
Ar  & $3.5\times10^{-12}$ & \dotfill & \dotfill & Sneep et al. (2005)\\
$\mathrm{H_2O}$ & $2.7\times10^{-12}$  & \dotfill & \dotfill & Tarafdar et al. (1973)\\
OH & $4.5\times10^{-13}$  & \dotfill & \dotfill & Sneep et al. (2005)\\
$\mathrm{CO_2}$ & $9.9\times10^{-12}$ & \dotfill & \dotfill & Sneep et al. (2005)\\
\hline
\end{tabular}
\label{table1}
\end{table}

\subsubsection{Resonance emission}

Similar as on Earth, the solar radiation pumps the atoms or molecules in the exosphere to
high energy levels. The excited atom or molecule then radiates a photon at a particular wavelength through
spontaneous emission. In the NUV bandpass of LUT, strong resonance emission transitions
mainly occur in the following emission lines: NaI$\lambda$330.33, 330.39, CaI$\lambda$272.25,
MgI$\lambda$285.30, MgII$\lambda\lambda$279.64, 280.35, AlI$\lambda$309.5,
and the OH($0-0$)($\mathrm{A^2\Sigma^+-X^2\Pi}$) band at wavelengths around 310~nm.

The intensity of each solar resonant scattering line is quantified for optically thin gas as
$4\pi I_s=gnl$. Here, $n$ and $l$ are the
gas concentration and line-of-sight path length, respectively. The solar-forced g-factor $g$
is defined as an emission probability per atom in units of $\mathrm{photon\ s^{-1}\ atom^{-1}}$.
The g-factor is determined by summing the probabilities of all transitions from multiple-states whose
population partitions are solved from the detailed equilibrium of every state that is usually
not in thermodynamic equilibrium. Assuming the vertical density distribution prescribed by
hydrostatic equilibrium, the sky brightness contributed by resonance emission can be
predicted by summing all possible resonance fluorescence lines within the NUV bandpass of LUT,
  \begin{equation}
    \label{eq:1}
 {I_s=\frac{1}{4\pi}\sum gn_0\Delta H
    }
  \end{equation}
where $n_0$ and $\Delta H$ are the surface concentration and scale height for each constituent
in the lunar exosphere, respectively.

In order to provide a strong constraint on the resonance emission, a universal scale height
$\Delta H=100$km is adopted in the subsequent calculations. We compile the g-factors from the literature.
The g-factors given in Killen et al. (2009) are calculated for the atmosphere of Venus
because Venus' atmosphere is now best understood among those outside Earth in the solar system.
These values are transformed to that of the Moon according to the relationship $g\propto r^{-2}$,
where $r$ is the distance from the Sun.
The g-factors depend not only on the distance from the Sun, but also on the Doppler velocity
$\upsilon_r$ of the Moon relative to the Sun (i.e., the Swing effect).
When $\upsilon_r=0$ the g-factors are minimal because of the Fraunhofer lines in
the solar spectrum. These minimum g-factors are adopted in our calculations
since the average radial velocity of the Moon is only $\upsilon_r <1.5\ \mathrm{km\ s^{-1}}$.
The uncertainty caused by the radial velocity is only a few percent for the values of the g-factors
(Stubbs et al. 2010).
Table 4 lists the adopted g-factors (at $r$=0.47~AU except for the g-factor of the OH radical at 1~AU),
and the corresponding surface concentrations. All the surface concentrations, except for the OH radical,
are quoted from Wurz et al. (2007). Note that only upper limits are reported for surface
concentrations, except for Na. The OH concentration listed in
Table 4 is the upper limit inferred from the results of CHACE (see Section 3). The resonance emission 
appears to be dominated by
the OH($0-0$)($\mathrm{A^2\Sigma^+-X^2\Pi}$) band emission at 308.5nm.
The sky brightness contributed by the resonance emission is predicted to be
$I_s<8.6\ \mathrm{ photon\ cm^{-2}\ s^{-1}\ arcsec^{-2}}$
according to Eq (3). This brightness is fainter than that of a point source with AB=13~mag.

\begin{table}
\small
\caption{Solar-forced g-values and corresponding surface concentrations for
strong resonant emission lines in the lunar exosphere. }
\begin{tabular}{lllll}
\hline
Species & Wavelength & $g$ & $n$ & References \\
 & nm & $\mathrm{photon\ s^{-1}\ atom^{-1}}$ & $\mathrm{cm^{-3}}$ & \\
\hline
NaI & 330.33 & $2.45\times10^{-3}$ & 75 & Killen et al. (2009)\\
    & 330.39 & $2.65\times10^{-3}$ &    & Killen et al. (2009)\\
CaI & 272.25 & $5.91\times10^{-3}$ & $<1$ & Killen et al. (2009)\\
MgI & 285.30 & $1.50\times10^{-1}$ & $<6000$ & Killen et al. (2009)\\
MgII& 279.64 & $3.69\times10^{-1}$ &         & Killen et al. (2009)\\
    & 280.35 & $1.71\times10^{-1}$ &         & Killen et al. (2009)\\
SiI & 252.60 & $8.15\times10^{-3}$ & $<48$   & Morgan et al. (1997)\\
AlI & 309.20 & $1.72\times10^{-1}$ & $<55$   & Morgan et al. (1997)\\
OH(0-0) & $\langle308.7\rangle$ & $1.04\times10^{-3}$ &  $<2\times10^9$ & Killen et al. (2009)\\
   &  &  &   & Feldman et al. (2010)\\
   &  &  &   & Schleicher et al. (1988)\\
\hline
\end{tabular}
\label{table1}
\end{table}

\subsubsection{Emissive photodissociation}

The molecules in the exosphere could be directly destroyed by solar ultraviolet emission
through the photodissociation process. The most likely reaction occurring is
the photodissociation of $\mathrm{H_2O}$. The reaction results in an
excited OH radical in the $\mathrm{A^2\Sigma^+}$ state, and then is followed by an
OH($0-0$)($\mathrm{A^2\Sigma^+-X^2\Pi}$) transition (e.g., Bertaux 1986):
%  \begin{eqnarray}
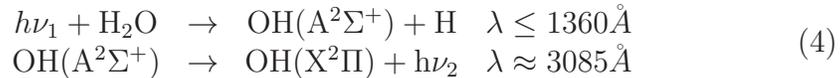
\begin{equation}
    \label{eq:1}
\begin{array}{cccc}
 h\nu_1+\mathrm{H_2O} & \rightarrow & \mathrm{OH(A^2\Sigma^+)+H} &  \lambda\leq1360\AA \\
  \mathrm{OH(A^2\Sigma^+)}& \rightarrow & \mathrm{OH(X^2\Pi)+h\nu_2} & \lambda\approx3085\AA
\end{array}
\end{equation}
%  \end{eqnarray}
On the Moon at 1~AU from the Sun, the ``excitation rate'' for the two reactions is
$P\sim5.48\times10^{-7}\ \mathrm{s^{-1}}$ (Crovisier 1989). The sky brightness contributed by the photodissociation
is predicted by the equation
\begin{equation}
I_c=\frac{1}{4\pi}n_0P\Delta H
\end{equation}
where
$n_0=6\times10^9\ \mathrm{cm^{-3}}$ is the upper limit to the surface concentration of $\mathrm{H_2O}$,
inferred from the results of CHACE.
By assuming a scale height $\Delta H=100$~km,
Eq. (5) predicts an upper limit of
$I_c=7.0\times10^{-2}\ \mathrm{ photon\ cm^{-2}\ s^{-1}\ arcsec^{-2}}$
for the sky background brightness caused by the OH photodissociation.

\subsubsection{Total sky background}

The total sky brightness is the sum of individual contributions
$I_b=I_p+I_s+I_c\approx I_s+I_c <8.7\ \mathrm{ photon\ cm^{-2}\ s^{-1}\ arcsec^{-2}}$,
where the contribution of Rayleigh scattering is excluded, 
because this emission is negligible compared with the other two processes.

\section{Count Rates and Signal-to-noise Ratio}

\subsection{Count rates}

The count rates expected for the LUT detector are calculated in this section for both object and sky background.
We refer the readers to Table 1 for the definition of the parameters used in the
subsequent equations.

The count rate within $3\times3$ pixels is calculated for a point source as follows
 \begin{equation}
    \label{eq:1}
 {R_*=\frac{1}{4}N_*\pi d^2\eta_{\mathrm{opt}}f_p\overline{QE}\ \mathrm{e^-\ s^{-1}}
    }
  \end{equation}
where $N_*$ is the photon flux of the point-source in units of $\mathrm{photon\ s^{-1}\ cm^{-2}}$.
A count rate of $R_*\approx58.3\ \mathrm{e^-\ s^{-1}}$ (within $3\times3$ pixels) is predicted
by inserting the parameter values of Table 1 and the value of $N_*$ estimated in Section 4.1
into the above equation.

The sky background count-rate per pixel also depends on the solid angle subtended by
each pixel $\Delta\Omega$, and is calculated by
\begin{equation}
    \label{eq:1}
{R_s=\frac{1}{4}\pi d^2\eta_{\mathrm{opt}}\Delta\Omega I_b\overline{QE}\ \mathrm{e^-\ s^{-1}\ pixel^{-1}}
}
\end{equation}
The solid angle per pixel is determined to be $\Delta\Omega=(\alpha d_p)^2=22.7\ \mathrm{arcsec^2\ pixel^{-1}}$
for LUT, where $\alpha=206265\symbol{125}/f$ and $f$ is the focal length.
An upper limit of $1.2\times10^3\ \mathrm{e^-\ s^{-1}\ pixel^{-1}}$ is predicted for the count rate per pixel
by substituting the total sky brightness predicted in Section 4.2.4 into Eq. (7).

\subsection{Signal-to-noise ratio}

The S/N ratio pertaining to a point source observation with exposure time $t$ (an
exposure time of 30 seconds is adopted in the subsequent calculations)
is approximately determined
by the traditional ``CCD'' equation (Merline \& Howell 1995):
 \begin{equation}
    \label{eq:1}
{\frac{S}{N}\simeq\frac{R_*t}{\sqrt{R_*t+n_{\mathrm{pix}}\bigg(1+\frac{n_{\mathrm{pix}}}{n_\mathrm B}
\bigg)(R_st+R_bt+Dt+RN^2+G^2\sigma^2_f})}
}
\end{equation}
Here, $R_b$ is the count rate per pixel due to stray light caused by the telescope. A stray light
simulation indicates that $R_b$ is as high as
$22\ \mathrm{e^-\ s^{-1}\ pixel^{-1}}$ in the worst case, and is significantly reduced to
$1\ \mathrm{e^-\ s^{-1}\ pixel^{-1}}$ in the best case (Cao et al. 2010). The stray light level
mainly depends on the Sun's elevation.
When $G$ is uniformly
distributed in (-1/2, 1/2), the variance caused by the A/D converters (i.e., the digitization noise) is
$\sigma^2_f=\int_{-1/2}^{1/2} f^2 df=0.289$. The parameter
$n_\mathrm{pix}=3\times3$ is the number of pixels in the aperture,
and $n_\mathrm B$ the number of pixels used for the background determination.
More the number of background pixels used,
better the background correction and digitization (and the higher the S/N ratio).
In astronomical observations, we typically have $n_\mathrm{pix}/n_\mathrm B\ll1$.
By ignoring the small elements in Eq (8), the S/N ratio can instead be estimated by the
simplified ``CCD'' equation (Mortara \& Fowier 1981; Gullixson 1992 and cf. NOAO/KPNO CCD instrument manuals)
\begin{equation}
    \label{eq:1}
{\frac{S}{N}\approx\frac{R_*t}{\sqrt{R_*t+n_{\mathrm{pix}}(R_st+R_bt+Dt+RN^2)}}
}
\end{equation}

By using the predictions of the above sections and the CCD performance parameters listed in Table 1,
we calculate the S/N ratio for an observation of a point source with AB=13~mag using the above equation.
In the worst case with the strongest stray light, the S/N ratio is predicted to be
$\sim19$ when the emission from the sky background is ignored.
The S/N ratio is reduced to $\sim3$
when the upper limit to the sky brightness is inserted.
This result means that the potential effect of the lunar exosphere on
LUT performance may be considerable if the exosphere is as dense as that suggested by CHACE.

\section{Discussions}

How much OH occurs in the lunar exosphere is a very important issue because
the emission from the OH($0-0$)($\mathrm{A^2\Sigma^+-X^2\Pi}$) transitions may be
(the dominant) component of the sky background brightness; also this emission is
commonly used as a
tracer for $\mathrm{H_2O}$ molecules. So far, there are only two in-situ experiments
that have measured the chemical abundances in the lunar exosphere.
Surprisingly, the OH concentration recently reported by CHACE is much higher
than that previously reported by LACE, by eight orders of magnitude. Here, we briefly summarize the existing
arguments (both observation and theory) on the OH radical issue, and
determine the S/N ratios for point source observations at different OH concentrations.

\subsection{HST observations of the Moon limb}

Stern et al. (1997) observed the lunar atmosphere using the HST
Faint Object Spectrograph and High resolution Spectrograph in the mid-UV band.
Because of the HST bright objects constraint, observations are not permitted
closer than 1.2~$\mathrm{R_M}$ from the lunar limb.
The authors did not detect OH($0-0$)($\mathrm{A^2\Sigma^+-X^2\Pi}$)
emission lines resulting in a $5\sigma$ upper limit
of $\sim10^6\ \mathrm{molecules\ cm^{-3}}$ for the OH surface concentration (Wurz et al. 2007).

\subsection{Ion sputtering}

Solar wind protons can penetrate
the lunar surface material to a depth of 0.05 to 0.1$\mu$m.
The protons then react with the lunar material to form chemically
adsorbed $\mathrm{H_2O/OH}$ molecules (e.g., Stern 1999;
Starukhina \& Shkuratov 2000; Arnold 1979).
These chemically adsorbed $\mathrm{H_2O/OH}$ molecules are stable below
a temperature $\sim500$~K (Hibbitts et al. 2010; Dyar et al. 2010).
Sputtering caused by high energy protons from the solar wind seems
a reasonable mechanism to produce water vapor and gaseous OH radicals in the
exosphere (Morgan et al. 1997; Wurz et al. 2007; Killen et al. 1999; Killen \& Ip  1999;
Hunten \& Sprague 1997; Johnson \& Baragiola 1991)\rm.
On the lunar surface, the proton flux of the solar wind is $j\sim10^8\ \mathrm{p^+\ cm^{-2}\ s^{-1}}$.
The sputtering results in a surface concentration of $\mathrm{H_2O/OH}$ molecules estimated as
$n\sim j\eta\upsilon^{-1}$, where $\eta\sim0.1$ is the production rate per
proton (e.g., Crider et al. 2002), and
$\upsilon\sim10^2\ \mathrm{m\ s^{-1}}$ is the typical velocity of a particle.
This yields an OH surface concentration of $\sim10^3\ \mathrm{molecules\ cm^{-3}}$.

The balance between ion sputtering and photodissociation caused by solar ultraviolet
photons indicates a column density of $N=1.8\times10^{10}\ \mathrm{cm^{-2}}$
for the resident OH radicals, corresponding to a limb brightness of
50\,Rayleigh (Morgan et al. 1991).
Assuming a scale height of $\Delta H=100$~km, the column density results in a
surface concentration of $\sim10^3\ \mathrm{cm^{-3}}$ for the OH radicals.

\subsection{OH escape}

Several space missions recently reported the chemically adsorbed $\mathrm{H_2O/OH}$ molecules
on the lunar surface. The adsorbed $\mathrm{H_2O/OH}$ molecules are identified through their 3~$\mu$m absorption features
(Clark et al. 2009; Pieters et al. 2009; Sunshine et al. 2009).
Sunshine et al. (2009) reported a variation of the 3~$\mu$m absorption features:
the absorption depth decreases with the Sun's elevation. This
implies a variation of the $\mathrm{H_2O/OH}$ column density of
$\Delta N\approx8\times10^{16}\ \mathrm{cm^{-2}}$ according to
the simulation performed by Starukhina et al. (2010).
Assuming that a fraction of the OH molecules escapes from the lunar surface in
half of a lunar day, the inferred concentration
in the exosphere is $n\sim\Delta N/\upsilon t\sim 10^7\ \mathrm{molecules\ cm^{-3}}$,
where $\upsilon\sim10^2\ \mathrm{m\ s^{-1}}$ is the typical velocity of a particle.
It requires a time scale of $t'\sim\Delta N/j\sim100$\,year
to recover the lost hydrogen through the ion sputtering, where
$j\sim10^8 \mathrm{p^+\ cm^{-2}\ s^{-1}}$ is the solar wind proton flux.
This shows that the lost hydrogen cannot be restored until the next lunar morning.
Starukhina et al. (2010) therefore argued that the reported variations are probably caused by additional
thermal emission from the illuminated lunar surface, while the emissivity was fixed
to one for all the wavelengths in the removal of the thermal continuum from the reflected spectra.
Following the above argument, we can estimate that the OH surface concentration is lower than
$\sim10^3\ \mathrm{molecules\ cm^{-3}}$ by assuming the lost hydrogen could be restored in
a time scale of $t'\sim10^6$s.

\subsection{Orbit decay of Chandrayaan-1}

If the lunar exosphere has a significant gas concentration at relatively low altitudes, we expect
that a spacecraft with a low lunar orbit experiences significant orbit decay due to ``atmospheric''
drag. However, the authors cannot find any report on orbit decay of
the Indian Chandrayaan-1 spacecraft during its flight in a 100~km orbit between 2008 November 12
and 2009 May 19. From this and the exospheric mass spectrum measured by the MIP CHACE instrument
of  Chandrayaan-I, a rough upper limit may be estimated for the lunar exospheric OH concentration.

Let us assume that the accumulated decay, if any, of Chandrayaan-1 during
its 177-day life must be less than 10~km to be insignificant. There have been about
2160 orbits with an orbital period of 2~hours. In other words, the average orbit decay per
orbit $\Delta h$ must satisfy $\Delta h/(R_{\rm m}+h)<2\times 10^{-6}$, where
$R_{\rm m}=1740$~km is the radius of the Moon and $h=100$~km is the orbit altitude.

For a near-circular orbit, according to the atmospheric drag theory, we have $\Delta h/(R_{\rm m}+h)
\approx \frac{2}{3}\cdot 3\pi\cdot\rho\cdot(R_{\rm m}+h)\cdot\frac{S\cdot C_D}{M}$, where $\rho$
is the ``atmospheric'' mass density. The mass of  Chandrayaan-1 is about 523~kg after the MIP
was released in 2008 February, its projected surface area $S$ is approximated as $1.5$~m$\times 1.5$~m,
and a typical value of 2 is chosen for the drag coefficient $C_D$. Substituting these parameters into
the above equation yields an upper limit to the total mass density of $\rho<2.6\times10^{-14}\ \mathrm{g\ cm^{-3}}$
for the exosphere at an altitude of $\sim100$km.

The mass spectrum measured by CHACE at an altitude of 96.6~km shows that the lunar exosphere is
dominated by H$_2$O/OH and by CO$_2$. The measured partial pressures are
$P_{\rm H_{2}O+OH}\sim 1\times 10^{-7}$~torr and
$P_{\rm CO_2}\sim 6\times 10^{-8}$~torr, which results in a concentration ratio of
$n_{\mathrm{OH}}:n_{\mathrm{H_2O}}:n_{\mathrm{CO_2}}=1:4:3$ assuming an identical
temperature for the three species. With the upper limit to $\rho$ deduced from the lack of orbit decay,
an upper limit to the OH concentration of
$n_{\mathrm{OH}}<7\times10^7\ \mathrm{cm^{-3}}$ can be derived from the equation $\rho=\sum n\mu/N_A$,
where $\mu$ is the molecular weight of each species and
$N_A=6.02\times10^{23}\ \mathrm{mol^{-1}}$ is the Avogadro number. This value is lower than that inferred in
Section 4.2.1 by one order of magnitude.

%By defining an effective ``temperature'' $T$, the
%partial pressure $P_i$ can be related to the partial concentration $n_i$ through the ideal gas law
%$P_i=n_i kT$, where $k$ is the Boltzmann constant. According to the CHACE mass spectrum, the ideal gas law means
%$n_{\rm OH}\approx 1/5\cdot n_{\rm H_{2}O+OH}$.
%With the upper limit of $\rho$ that is deduced by our orbit decay,
%we have $n_{\rm OH}<7.5\times 10^8$~cm$^{-3}$ for the lunar exosphere at roughly an altitude of 100~km,
%which is consistent with the value inferred in Section 4.2.1.

%Neglecting other mass components and the small molecular mass difference between H$_2$O and OH, we have a total
%``atmospheric'' mass density of $\rho\sim 2.4\times 10^{-13}(300~{\rm K}/T)$~g~cm$^{-3}$.
%One immediately obtains $T>2400$~K from the relation by adopting the upper limit of $\rho$.
%The effective ``temperature'' is found to be roughly consistent with the temperature estimated in Section 4.2.1.

\subsection{Signal-to-noise ratios for different OH concentrations}

We predict the sky background brightness and
the corresponding count rates within $3\times3$ pixels
for different OH surface concentration levels, ranging from $1\times10^4$ to $2\times10^9\ \mathrm{molecules\ cm^{-3}}$.
The results are listed in Table 5 and shown in Figure 1. 
The curve is fairly steep at the high concentration end and,
rather shallow at the low concentration end.
The turnover of the curve occurs at
$n\sim10^{6-7}\ \mathrm{molecules\ cm^{-3}}$, because the emission of the sky background begins to be
dominated by the H($0-0$)($\mathrm{A^2\Sigma^+-X^2\Pi}$) transitions.

At a concentration of
$2\times10^8\ \mathrm{molecules\ cm^{-3}}$, the S/N ratio of an observation of a given AB=13~mag point source
will be above 8, although
the LUT detection performance can be degraded by the existence of the OH(0-0) emission.
The effect of a ``bright'' OH exosphere on LUT's performance cannot be ignored in the worst case scenario.

In-situ measurements are essential to
understand the
origin of the $\mathrm{H_2O/OH}$ molecules on the Moon. We think that a
lunar-based telescope operating
in the NUV band will be able to provide significant constraints on the $\mathrm{H_2O/OH}$ component by
examining the sky background.

\begin{table}
\small
\caption{Sky background brightness and corresponding
count rates at different OH concentration levels.}
\begin{tabular}{lll}
\hline
OH concentration & Sky background brightness & Count-rates \\
 & & within $3\times3$ pixels\\
 $\mathrm{molecules\ cm^{-3}}$ & $\mathrm{photon\ cm^{-2}\ s^{-1}\ arcsec^{-1}}$ & $\mathrm{e^-\ s^{-1}}$ \\
\hline
$1\times10^4$ & $2.5\times10^{-2}$ & 33.0 \\
$1\times10^5$ & $2.6\times10^{-2}$ & 33.5 \\
$1\times10^6$ & $3.0\times10^{-2}$ & 38.6 \\
$1\times10^7$ & $6.8\times10^{-2}$ & 89.0 \\
$1\times10^8$ & $4.6\times10^{-1}$ & 592.7\\
$2\times10^8$ & $8.9\times10^{-1}$ &1,152.4\\
$2\times10^9$ & $8.7\times10^{0}$ & 11,214.0\\
\hline
\end{tabular}
\label{table1}
\end{table}

\begin{figure}
\label{figure1}
\begin{center}
\includegraphics*[width=10cm,angle=0]{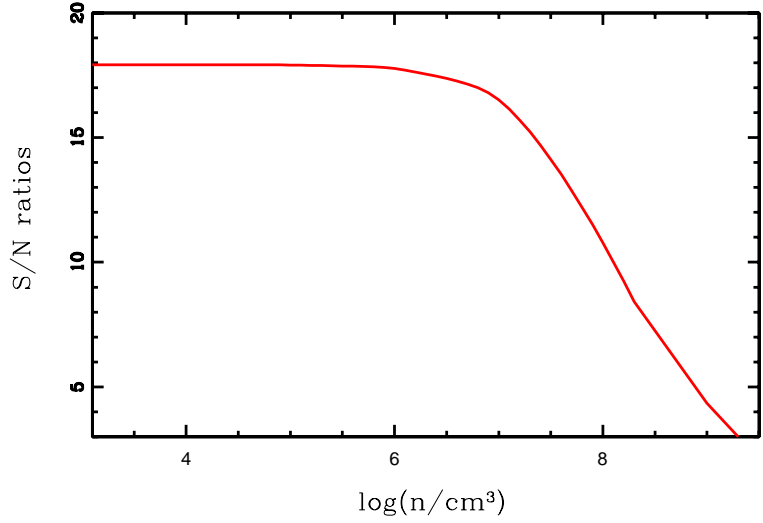}
\end{center}
\caption{Signal-to-noise ratios pertaining to an AB=13~mag point source versus the
surface OH concentrations.}
\end{figure}

\section{Conclusions}

We estimate the near-ultraviolet sky background brightness of the lunar exosphere. 
Our calculations show that the sky brightness
is $<8.7\ \mathrm{photon\ s^{-1}\ cm^{-2}\
arcsec^{-2}}$ within the wavelength range 245-340~nm.
The signal-to-noise analysis indicates that the detection performance of LUT can be
degraded by the sky background emission in most cases. An AB=13~mag point source can
be detected by LUT
at a signal-to-noise ratio above 8 when the OH concentration is less than
$2\times10^8\ \mathrm{molecules\ cm^{-3}}$. However, the effect on the
performance is not ignorable when the exosphere is as dense as suggested
by CHACE.

\section*{Acknowledgments}
The authors thank the anonymous referee for his/her comments that improve the quality of the
paper. Special thanks go to Paul Wesselius, the scientific editor, for his/her great effort in 
improving the manuscript.
The study is funded by the National Science and Technology Major Project. JW is supported by the
National Science Foundation of China (under grant 10803008). LC is supported by the
National Science Foundation of China (Grant Nos. 10978020 and 10878019). JW, JSD, YLQ and
JYW are supported by the National Basic Research Program of China (Grant 2009CB824800).
The authors would like to thank F. Tian, H. Zhao and Prof. J. Y. Hu for
valuable discussions and suggestions. We would like to thank James Wicker for help with the language.

%\section{acknowledgements}

%\section{Citations}
%\label{Section 3}

%\begin{itemize}
%\item Parenthetical: \verb|\citep{WB96}| produces \citep{WB96}.
%\item Textual: \verb|\citet{WB96}| produces \citet{WB96}.
%\item An affix and part of a reference:
%   \verb|\citep[e.g.][Ch. 2]{WB96}|
%   produces \citep[e.g.][Ch. 2]{WB96}.
%\end{itemize}

%%%%%%%%%%%%%%%%%%%%%%%%%%%%%%%%%%%%%%%%%%%%%%%%%%%%%%%%%%%%%%%%%%%%%%%%%%%%%
%% Appendices
% \appendix

%\clearpage

\end{document}